\documentclass[nonacm,sigplan]{acmart}

\usepackage[]{hyperref}

\usepackage{tikz}
\usepackage{amsmath}

\usepackage{amssymb}

\usepackage{booktabs}
\usepackage{xspace}
\usepackage{mathtools}
\usepackage[ruled, vlined]{algorithm2e}
\usepackage{enumitem}
\usepackage{subcaption}
\usepackage{svg}
\usepackage{graphicx}
\usepackage{bbding}
\usepackage{multirow}
\usepackage{amsmath}
\usepackage{breqn}
\usepackage{algorithmic}

\usepackage{soul}
\usepackage{makecell}
\usepackage{balance}

\newcommand{\name}{\text{FlexSP}\xspace}

\DeclareMathOperator*{\argmin}{arg\,min}

\newif\ifasplos

\ifasplos
\copyrightyear{2025}
\acmYear{2025}
\setcopyright{acmlicensed}\acmConference[ASPLOS '25]{Proceedings of the 30th ACM International Conference on Architectural Support for Programming Languages and Operating Systems, Volume 2}{March 30-April 3, 2025}{Rotterdam, Netherlands}
\acmBooktitle{Proceedings of the 30th ACM International Conference on Architectural Support for Programming Languages and Operating Systems, Volume 2 (ASPLOS '25), March 30-April 3, 2025, Rotterdam, Netherlands}
\acmDOI{10.1145/3676641.3715998}
\acmISBN{979-8-4007-1079-7/2025/03}
\fi

\begin{document}

\title{\name: Accelerating Large Language Model Training via Flexible Sequence Parallelism}


\author{Yujie Wang}
\authornote{School of Computer Science \& Key Lab of High Confidence Software Technologies (MOE), Peking University}
\affiliation{
  \institution{Peking University}
  \city{Beijing}
  \country{China}
}
\email{alfredwang@pku.edu.cn}

\author{Shiju Wang}
\affiliation{
  \institution{Beihang University}
  \city{Beijing}
  \country{China}
}
\email{21373455@buaa.edu.cn}

\author{Shenhan Zhu}
\authornotemark[1]
\affiliation{
  \institution{Peking University}
  \city{Beijing}
  \country{China}
}
\email{shenhan.zhu@pku.edu.cn}

\author{Fangcheng Fu}
\authornotemark[1]
\affiliation{
  \institution{Peking University}
  \city{Beijing}
  \country{China}
}
\email{ccchengff@pku.edu.cn}

\author{Xinyi Liu}
\authornotemark[1]
\affiliation{
  \institution{Peking University}
  \city{Beijing}
  \country{China}
}
\email{xy.liu@stu.pku.edu.cn}

\author{Xuefeng Xiao}
\affiliation{
  \institution{ByteDance Inc.}
  \city{Beijing}
  \country{China}
}
\email{xiaoxuefeng.ailab@bytedance.com}

\author{Huixia Li}
\affiliation{
  \institution{ByteDance Inc.}
  \city{Beijing}
  \country{China}
}
\email{lihuixia@bytedance.com}

\author{Jiashi Li}
\affiliation{
  \institution{ByteDance Inc.}
  \city{Shenzhen}
  \country{China}
}
\email{lijiashi@bytedance.com}

\author{Faming Wu}
\affiliation{
  \institution{ByteDance Inc.}
  \city{Beijing}
  \country{China}
}
\email{wufaming@bytedance.com}

\author{Bin Cui}
\authornotemark[1]
\authornote{Institute of Computational Social Science, Peking University (Qingdao)}
\affiliation{
  \institution{Peking University}
  \city{Beijing}
  \country{China}
}
\email{bin.cui@pku.edu.cn}

\renewcommand{\shortauthors}{Yujie Wang et al.}

\begin{abstract}



Extending the context length (i.e., the maximum supported sequence length) of LLMs is of paramount significance. 
To facilitate long context training of LLMs, sequence parallelism has emerged as an essential technique, which scatters each input sequence across multiple devices and necessitates communication to process the sequence. 
In essence, existing sequence parallelism methods assume homogeneous sequence lengths (i.e., all input sequences are equal in length) and therefore leverages a single, static scattering strategy for all input sequences. However, in reality, the sequence lengths in LLM training corpora exhibit substantial variability, often following a long-tail distribution, which leads to workload heterogeneity. 

In this paper, we show that employing a single, static strategy results in inefficiency and resource under-utilization, highlighting the need for adaptive approaches to handle the heterogeneous workloads across sequences. 
To address this, we propose a heterogeneity-adaptive sequence parallelism method. 
For each training step, our approach captures the variability in sequence lengths and assigns the optimal combination of scattering strategies based on workload characteristics. 
We model this problem as a linear programming optimization and design an efficient and effective solver to find the optimal solution. 
Furthermore, we implement our method in a high-performance system that supports adaptive parallelization in distributed LLM training.
Experimental results demonstrate that our system outperforms state-of-the-art training frameworks by up to 1.98$\times$.
Source code is available at \url{https://github.com/PKU-DAIR/Hetu-Galvatron}.

\end{abstract}

\maketitle 
\ifasplos
\else
\pagestyle{plain} 
\fi

\section{Introduction} \label{sec:intro}  

Large Transformer models~\cite{DBLP:conf/nips/transformer,DBLP:conf/naacl/BERT,gpt1,DBLP:journals/chinaf/ShaoGLDYYZBQ24,DBLP:journals/dase/LiDWHCXDL23}, represented by Large Language Models (LLMs)~\cite{gpt2,DBLP:gpt3,gpt4,DBLP:googleT5,DBLP:llama,DBLP:llama2,DBLP:OPT,DBLP:journals/dase/ZhouSL24,zhang2024deep}, have made astonishing achievements in the field of artificial intelligence (AI). 
Recently, there is an increasing need to extend the context length of LLMs, which represents the maximum supported sequence length of the LLMs ~\cite{ yang2023baichuan2openlargescale, deepseekai2024deepseekv2strongeconomicalefficient, dubey2024llama3herdmodels}. 
Therefore, long-context LLM training has garnered extensive attention from both the academia and industry ~\cite{DBLP:RSA,DBLP:ring-attn,distflashattn,DBLP:striped_attn, DBLP:journals/corr/abs-2004-05150, DBLP:conf/acl/DaiYYCLS19,bigbird}. 

It is well known that training LLMs with longer context lengths demands increasingly more device (e.g., GPU) memory, and a promising paradigm is to parallelize the training inputs over multiple devices. 
Particularly, sequence parallelism (SP)~\cite{DBLP:deepspeed-sp-Ulysses,DBLP:megatron-sp,DBLP:RSA,DBLP:ring-attn,distflashattn,DBLP:striped_attn} has emerged as a pivotal technique for long-context LLM training. 
In a nutshell, SP splits each training input in the sequence dimension and scatters different shards onto multiple devices to amortize the memory consumption. 
In order to carry out the training, it necessitates communication among the devices to exchange the intermediate results during the forward and backward propagation. 
By nature, if we wish to support longer training inputs, we shall increase the SP degree\footnote{For simplicity, we use the abbreviation ``SP degree'' to denote the parallelism degree of SP. This terminology extends naturally to other parallelism strategies (e.g., data parallelism degree).} (usually accompanied by a decrease in the data parallelism degree) to avoid encountering out-of-memory errors, while the communication overhead increases in the meantime, leading to efficiency degradation. 

Although SP delivers a method to support long-context LLM training, existing systems assume the training inputs are homogeneous in terms of their lengths and apply the same SP degree to all data parallel model replicas, yet such a fixed-length assumption does not hold in practice. 
Typically, the training samples of LLMs are usually organized as sequences of tokens, and a training corpus consists of varied-length sequences. 
In order to address the discrepancy between the fixed-length assumption and the varied-length characteristic, sequence packing is a widely used data pre-processing technique. 
Specifically, denote $c$ as the maximum number of tokens supported for each model replica. Sequence packing concatenates multiple sequences into a longer one and ensure that the length of the packed sequence does not exceed $c$. (A sequence will be truncated if it exceeds $c$ by itself.)
Meanwhile, the attention masks and position indices are adjusted accordingly to avoid cross-contamination among the sequences. 
By this means, the training inputs\footnote{To avoid ambiguity, by saying ``training inputs'', we always mean the packed sequences that are fed into the training process.} can be (nearly) homogeneous in terms of their lengths, and the model gradients computed over a packed input are identical to that computed over the original, unpacked sequences. 

\begin{figure}[t]
    \centering
    \includegraphics[width=1.\linewidth]{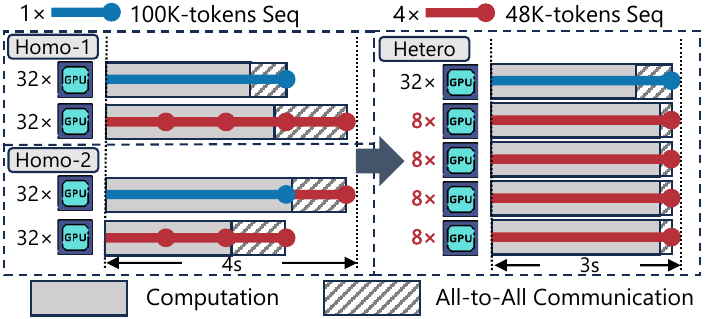}
    \caption{
        An example of heterogeneity-adaptive SP improving training efficiency for varied-length sequences.
    }
    \label{fig:obs}
\end{figure}

Nevertheless, we find that the aforementioned approach sacrifices the efficiency of short sequence processing to accommodate the memory requirement of long-sequence processing.
Fig.~\ref{fig:obs} showcases an example, where there are 64 devices in total and the context length of the training task is 192K. 
Using the aforementioned approach, it necessitates an SP degree of 32, resulting in two SP=32 groups. 
Assume the current training step needs to process five sequences with lengths of 100K, 48K, 48K, 48K, and 48K, respectively. 
Due to the homogeneous parallelism designs of existing works, no matter how we pack the sequences (Case Homo-1 or Homo-2 in left side of Fig.~\ref{fig:obs}), the short sequences (48K) must be processed with an SP degree of 32, and the All-to-All communication time and computation time of four 48K sequences are 1.2s and 2.8s, respectively.
However, if we allow for heterogeneous SP groups (i.e., allow groups to have non-unique SP degrees), we can form one SP=32 group to process the long sequence (100K) and four SP=8 groups to process the short sequences (48K) separately (Case Hetero in right side of Fig.~\ref{fig:obs}). 
By doing so, the computation time of four 48K sequences remains 2.8s, while the communication time decreases to only 0.2s,and the processing of the short sequences can be accelerated (from 4s to 3s in the example) due to the reduction in communication overhead. 
As we will detail in \S\ref{sec:obs}, common LLM training corpora follow long-tail distributions, i.e., there are very few long sequences while the short ones dominate the corpora. 
Consequently, by adjusting the SP groups, we can accelerate the training of most sequences, provisioning performance improvement. 

Inspired by this, this work develops \name, a Flexible Sequence Parallelism for the efficient training of LLMs over varied-length sequences. 
Unlike existing systems that organize training sequences to fit the homogeneous parallelism, \name adapts to the heterogeneous workloads of varied-length sequences by dynamically adjusting the parallelism.
Essentially, \name puts forward two key innovations:
\begin{itemize}[leftmargin=*]
\item \ul{\textit{Heterogeneous Sequence Parallel Groups:}}
As aforementioned, processing long sequences necessitates a higher SP degree to avoid out-of-memory errors, while processing short sequences has a better efficiency if we use a lower SP degree. 
Consequently, for each training step, \name adaptively forms multiple heterogenous SP groups so that sequences with diverse lengths can be processed with different groups, making a good tradeoff between memory consumption and training efficiency.

\item \ul{\textit{Time-Balanced Sequence Assignment:}}
To minimize the processing time of each sequence individually, a na\"ive approach is to assign each sequence to the smallest SP group that can handle it. 
However, due to the long-tail distribution of datasets, short sequences dominate, and such a na\"ive assignment causes small groups to handle too many workloads, causing a bottleneck. This leads to workload imbalances across the SP groups, where faster groups are forced to wait for slower ones in idle.
To deal with this problem, \name meticulously controls which SP group each sequence should be assigned to, striking a good balance across the heterogeneous SP groups.
\end{itemize}

\name co-optimizes these two factors adaptively according to the sequence length variation of each training step. 
Specifically, given the sequences with diverse lengths, we formulate a joint optimization problem to maximize training efficiency, which determines how to form the heterogeneous SP groups and how to assign each sequence to the most suitable group. 
To solve this problem, we transform it into a Mixed-Integer Linear Programming (MILP) problem by accurately modeling the computation cost, communication cost, and memory consumption of training over varied-length sequences. 
Subsequently, we decrease the complexity of the MILP problem via a sequence bucketing algorithm based on dynamic programming, making it efficient to solve for practical considerations. 

In addition, to resolve the potential issue that there are too many sequences that cannot be processed at once, we manage to chunk the sequences into micro-batches. 
In particular, we establish a series of propositions based on theoretical analysis and empirical observations. 
Based on this, a micro-batch chunking algorithm is devised, which aims to minimize the total training time of all micro-batches. 

We implement \name on top of PyTorch and conduct experiments with various model sizes, training datasets, and context lengths. 
Empirical results show that \name outperforms state-of-the-art LLM training systems by up to 1.98$\times$, demonstrating the effectiveness of flexible sequence parallelism in long-context LLM training. 

The contribution of this work are summarized as follows: 
(1) \ul{\textit{New System}}.
We introduce \name, a brand new distributed training LLM system for varied-length corpora. 
(2) \ul{\textit{New Perspective}}.
To the best of our knowledge, \name is the first to adaptively adjust the parallelism strategies given the diverse lengths, matching the heterogeneous workloads caused by varied-length sequences with heterogeneous parallelism. 
Existing homogeneous systems overlook the heterogeneous workloads of varied-length contexts, and typically pack sequences to similar lengths, sacrificing the efficiency of short sequences. 
In contrast, \name is based on first principles, supporting heterogeneous parallelisms that match the diverse workloads, 
proposing a new perspective of redesigning systems for more efficient training of powerful models (e.g., training over varied-length images/videos).
(3) \ul{\textit{State-of-the-art Performance}}.
Extensive experimental results verify that \name consistently achieves the best training efficiency compared to existing works.
























\section{Preliminary} \label{sec:preliminary}

\subsection{Parallelisms in Distributed Training}
As model sizes and data volumes grow rapidly, various distributed parallelism techniques are employed in LLM training to distribute the workload across multiple devices.

\subsubsection{Data Parallelism}
Data parallelism (DP)~\cite{DBLP:pytorch-ddp} splits the data along sample dimension. 
Each device is responsible for a part of the input samples, and the gradients need to be synchronized across the devices.
DP requires each device to maintain a complete copy of the model, which is redundant.
To address this, sharded data parallelism (SDP) has been proposed, such as DeepSpeed-ZeRO~\cite{DBLP:deepspeed-zero} and PyTorch FSDP~\cite{DBLP:pytorch-fsdp}. 
These methods not only split the data but also the model states across devices, allowing each device to store only a fraction of the model states and introducing additional communication to synchronize the model states.

\subsubsection{Sequence Parallelism}
Sequence parallelism (SP) also splits data, and can be considered a special form of DP. 
Unlike DP, which splits data across the sample dimension, sequence parallelism splits data across the sequence dimension. 
SP is designed to mitigate the memory shortage caused by increasingly longer context lengths of LLMs.
DeepSpeed-Ulysses~\cite{DBLP:deepspeed-sp-Ulysses} proposes Ulysses-style SP, which splits the sequence dimension of linear projection in MLP and attention module, dropout modules, and normalization modules, and employs All-to-All primitives to collect and distribute sequences.
{
\begin{align}
    \mathbf{Q}_s,\mathbf{K}_s,\mathbf{V}_s &= \mathbf{X}_s\mathbf{W}_Q,\mathbf{X}_s\mathbf{W}_K,\mathbf{X}_s\mathbf{W}_V \in \mathbb{R}^{\frac{N}{P}\times d} \label{eq:in_linear} \\
    \mathbf{Q}_h,\mathbf{K}_h,\mathbf{V}_h &= \text{AlltoAll}(\mathbf{Q}_s,\mathbf{K}_s,\mathbf{V}_s) \in \mathbb{R}^{N\times \frac{d}{P}} \label{eq:all2all} \\
    \mathbf{P}_h &= \text{softmax}\left(\frac{\mathbf{Q}_h\mathbf{K}_h^{\top}}{\sqrt{d}}\right)\mathbf{V}_h \in \mathbb{R}^{N\times \frac{d}{P}} \label{eq:attn} \\
    \mathbf{O}_s &= \text{AlltoAll}(\mathbf{P}_h\mathbf{W}_O) \in \mathbb{R}^{\frac{N}{P}\times d} \label{eq:out_linear}
\end{align}
}
In the attention module of Ulysses-style SP, 
each device holds a portion of sequences $\mathbf{X}_s \in \mathbb{R}^{\frac{N}{P}\times d}$ and the complete model parameters 
$\mathbf{W}_Q, \allowbreak \mathbf{W}_K, \allowbreak \mathbf{W}_V, \allowbreak \mathbf{W}_O \in \mathbb{R}^{d \times d}$
, where $N$ denotes the total sequence length, $P$ denotes the SP degree, and $d$ denotes the hidden size.
After the linear projection of query, key and value (Eq.~\ref{eq:in_linear}), three rounds of All-to-All communication are employed to collect the complete sequence on each device (Eq.~\ref{eq:all2all}).
The multi-head attention operation is then calculated on the complete sentence (Eq.~\ref{eq:attn}), and an another round of All-to-All communication is introduced to distribute the sequence across the devices (Eq.~\ref{eq:out_linear}).

Megatron-LM also proposes Megatron-style SP~\cite{DBLP:megatron-sp}, which splits only the dropout and normalization modules, and requires All-Gather and Reduce-Scatter communication. 
It can be treated a supplementary scheme proposed to be used in conjunction with Megatron-TP (\S\ref{other_parallel}), aimed at addressing the redundant activation memory usage of Megatron-TP, and they must have the same parallelism degree.

\subsubsection{Other parallelisms} \label{other_parallel}
\paragraph{Model Parallelism} 
Model parallelism distributes the model parameters across the cluster, which can be divided into two categories: tensor parallelism (TP) and pipeline parallelism (PP). 
TP splits the model vertically. 
Megatron-TP proposed by Megatron-LM~\cite{DBLP:megatron} is the most widely used, which splits the tensor multiplication computations in each attention layer and feed-forward layer across multiple devices, incorporating communication to synchronize the computation results. 
PP~\cite{DBLP:conf/nips/gpipe,DBLP:conf/sosp/pipedream,pipedream-flush,DBLP:journals/jcst/GuanLLWGL24} partitions the model horizontally, with GPipe~\cite{DBLP:conf/nips/gpipe}, PipeDream-Flush~\cite{pipedream-flush} being notable implementations. 
These approaches divide the model layers into multiple stages placed across different devices, pass intermediate computation results with point-to-point communication, and orchestrate the model execution into a pipeline. 

\paragraph{Context Parallelism}
Context parallelism (CP)~\cite{DBLP:RSA,DBLP:ring-attn,distflashattn,DBLP:striped_attn} also splits the sequence dimension. Compared to sequence parallelism (SP) which splits dropout and normalization module activations but necessitates complete sentence for attention operation, CP further distributes the attention operation.
Specifically, CP distributes sequence dimension of the query, key, and value across multiple devices, and involves additional ring communication to collect key and value for completing attention computations.
Such extra communication volume is substantial, thus CP allows the computation to overlap the extra communication overhead by conducting the ring communication and computation of attention operation chunk by chunk.

\subsubsection{LLM Training Systems and Hybrid parallelisms}
Modern LLM training systems~\cite{DBLP:megatron,Deepspeed} usually combine multiple dimensions of parallelisms and support hybrid parallelism.
For instance, Megatron-LM~\cite{DBLP:megatron} supports 4D parallelism, including DP, TP (along with Megatron-style SP), PP, CP. 
DeepSpeed~\cite{Deepspeed} supports SDP (DeepSpeed-ZeRO) and Ulysses-style SP.
However, all of these existing parallelism techniques and LLM training systems are designed for training corpora with homogeneous context lengths and employ single and static strategy, ignoring the variability of sequence lengths in real-world datasets.

\subsection{Common Techniques in LLMs Training}

\subsubsection{Gradient Accumulation} \label{pre:ga}
Gradient accumulation is a practical technique to train large batch of data under constrained memory capacity.
It simulates the large batch training by splitting data batch into several micro-batches, executing micro-batches sequentially, accumulating gradients and updating model until the last micro-batch finishes.

\subsubsection{Sequence Packing}
To train the varied-length sequences together, techniques such as padding or packing are often required. 
Padding involves extending or truncating all sequences into the same length, which introduces redundant computation and memory overhead due to the excess padding for short sequences. 
In contrast, sequence packing~\cite{krell2021efficient} is a more advanced and commonly used technique that concatenates sequences of different lengths into a single long sequence, where adjustments to the attention masks and position indices are required to prevent cross-contamination. 
Sequence packing eliminates the redundancy caused by padding the varied-length sequences.
\cite{icml-bestfitpack} proposes \textit{Best-fit Packing}, employing \textit{Best-Fit-Decreasing} (BFD) to pack sequences to best fit device memory. 
In our work, sequence packing is the default setting.

\iftrue
\begin{table}[!t]
    \centering
    \captionof{table}{
        Iteration time (s) of GPT-7B with SP, along with the All-to-All communication ratios of different sequence lengths (seq), batch size (bs) pairs and SP degrees,
        tested on 64 A100 GPUs, equipped with NVLink for intra-node (SP $\le$ 8) and InfiniBand for inter-node (SP $>$ 8) communication.
    }
    \label{tab:obs}
    \scalebox{.9}{
    \begin{tabular}{c|c|c|c|c|c}
    \toprule
    seq $\times$ bs  & SP=64         & SP=32          & SP=16          & SP=8           & SP=4           \\ \hline\hline
    4K $\times$ 1024 & \makecell{\underline{37.2}\\54.4\%}  & \makecell{\underline{33.6}\\45.0\%} & \makecell{\underline{27.9}\\33.2\%} & \makecell{\underline{19.2}\\8.1\%} & \makecell{\underline{\textbf{18.9}}\\\textbf{7.3\%}} \\ \hline
    8K $\times$ 512  & \makecell{\underline{37.6}\\51.9\%}  & \makecell{\underline{34.9}\\42.4\%} & \makecell{\underline{29.1}\\31.4\%} & \makecell{\underline{20.9}\\7.8\%} & \makecell{\underline{\textbf{20.4}}\\\textbf{7.1\%}} \\ \hline
    16K $\times$ 256  & \makecell{\underline{40.6}\\47.8\%}  & \makecell{\underline{37.6}\\38.6\%} & \makecell{\underline{32.6}\\29.2\%} & \makecell{\underline{23.4}\\6.9\%} & \makecell{\underline{\textbf{23.3}}\\\textbf{6.2\%}} \\ \hline
    32K $\times$ 128  & \makecell{\underline{47.5}\\41.6\%}  & \makecell{\underline{44.0}\\33.3\%} & \makecell{\underline{39.4}\\24.2\%} & \makecell{\underline{\textbf{30.4}}\\\textbf{5.7\%}} & OOM         \\ \hline
    64K $\times$ 64   & \makecell{\underline{61.5}\\34.1\%}  & \makecell{\underline{56.2}\\25.1\%} & \makecell{\underline{\textbf{51.8}}\\\textbf{18.6\%}} & OOM         & OOM         \\ \hline
    128K $\times$ 32   & \makecell{\underline{85.0}\\23.5\%}  & \makecell{\underline{\textbf{82.8}}\\\textbf{18.5\%}} & OOM         & OOM         & OOM         \\ \hline
    256K $\times$ 16   & \makecell{\underline{\textbf{137.2}}\\\textbf{16.4\%}} & OOM         & OOM         & OOM         & OOM         \\
    \bottomrule
    \end{tabular}
    }
\end{table}
\fi

\section{Observation and Opportunities} \label{sec:obs}
In this section, we introduce our observations, and illustrate the optimization opportunities.
\subsubsection*{Observation 1: Long sequences require large parallelism groups, which leads to inefficiency.}
As mentioned in \S\ref{sec:intro}, existing works simply assume a homogeneous sequence length and therefore apply a homogeneous SP degree in each training task. 
To better illustrate the drawbacks of homogeneous SP design, we conduct a small testbed to assess its efficiency when facing different levels of sequence lengths. 
Tab.~\ref{tab:obs} presents the end-to-end execution time along with the proportion of All-to-All communication. 
Particularly, for each row in Tab.~\ref{tab:obs}, we generate fixed-length sequences until the total token number has reached 4 million, and train these sequences with divergent SP degrees. 
Firstly, we find that long sequences require larger parallelism groups due to memory constraint.
For instance, a 32K sequence can be handled by SP groups with a degree of 8, whereas a 128K sequence requires a degree of at least 32 to fit within device memory.
Secondly, larger parallelism groups can result in inefficiencies, primarily because of increased communication overhead.
Smaller groups tend to perform better.
For example, with 512 sequences of 8K, an SP group with a degree > 8 leads to communication time exceeding 31.4\% (9.1s) due to the slow inter-node bandwidth.
In contrast, groups with a degree of 8 reduce communication to just 7.8\% (1.6s), benefiting from the high-bandwidth intra-node connection.
In this case, the computation time remains unchanged (approximately 19s) for variable-degree SP groups, as expected, and the communication is much more efficient on smaller parallelism groups.

\subsubsection*{Observation 2: Real-world datasets present skewness in sequence length distribution, exhibiting a long-tail distribution.}
\begin{figure}[!t]
    \centering
    \includegraphics[width=1.\linewidth]{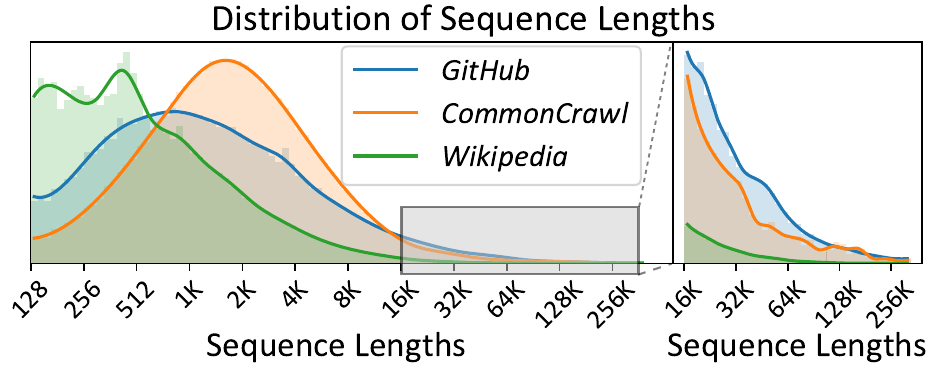}
    \captionof{figure}{
        Distribution of sequence lengths across different datasets.
        The height of each bar represents the percentage of sequences in the corresponding length range.
        Details of excessively long sequences are expanded into the right panel.
    }
    \label{fig:seq-dist}
\end{figure}

Fig. ~\ref{fig:seq-dist} illustrates the sequence length distribution for three popular LLM training datasets, which are \textit{GitHub}, \textit{CommonCrawl}, and \textit{Wikipedia}.
We observe that all three datasets exhibit a pronounced uni-modal long-tail distribution, with the majority of sequences falling below 8K in length, while only a small fraction of sequences exceed 32K.
\textit{GitHub} contains the largest number of excessively long sequences, followed by \textit{CommonCrawl}, with \textit{Wikipedia} having the fewest. 
Consequently, the sequence lengths in real-world datasets demonstrate a notable skewness, following a long-tailed distribution.

\subsubsection*{Optimization Opportunity: Matching heterogeneous parallelism groups with heterogeneous sequence lengths.}

Current systems are tailored for homogeneous sequence lengths and employ single, static parallelism strategies throughout the training process.
However, dealing with sequences of long-tail distribution in lengths requires large parallelism groups to accommodate the excessively long sequences,  which diminishes efficiency for shorter sequences.
For instance, when a sequence of 128K exists in the dataset, all sequences of 8K are forced to use SP groups of size at least 32, failing to enjoy the efficient smaller groups.
Furthermore, given that short sequences are more common in skewed distributions, this inefficiency is pronounced. 
Therefore, we propose to adapting appropriate parallelism strategies, crafting heterogeneous parallelism groups to match the heterogeneous workloads caused by varied-length sequences. 
Specifically, we wish to form small groups for short sequences to improve efficiency, while retaining large groups for long sequences to avoid out-of-memory errors.
Additionally, we need to properly control the assignment of sequences to balance the execution time among all parallelism groups. 
Such flexible, heterogeneity-adaptive strategies will improve communication efficacy and overall system performance.

\newcommand{\mbname}{parallelism planner\xspace}
\newcommand{\Mbname}{Parallelism Planner\xspace}
\newcommand{\gbname}{sequence blaster\xspace}
\newcommand{\Gbname}{Sequence Blaster\xspace}

\begin{figure}[!t]
    \centering
    \includegraphics[width=1.\linewidth]{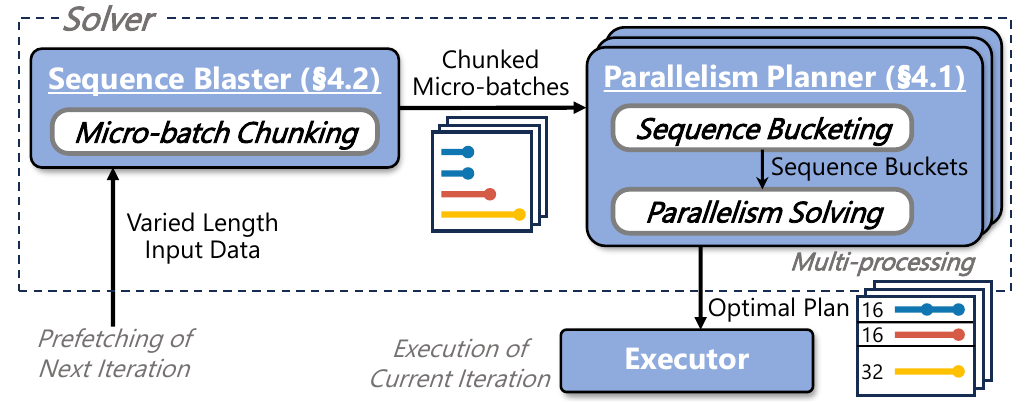}
    \caption{
        \name system overview.
    }
    \label{fig:overview}
\end{figure}

\section{\name} \label{sec:method}
Fig.~\ref{fig:overview} outlines the system overview of \name, which consists of the solver and the executor.
Given a batch of sequences with diverse lengths, the solver deduces the optimal plan of heterogeneous parallelism groups and sequence assignment. 
In particular, there are two major steps in the solver. 
Firstly, the \gbname chunks the sequences into micro-batches, ensuring that each micro-batch will not be too large to accommodate. 
Secondly, the \mbname is responsible for solving the optimal plan for each micro-batch to minimize its execution time. 
Following the optimal plan, the executor carries out the training of one iteration.

In this section, we focus on the details of \name's solver. 
Frequently used notations are listed in Tab.~\ref{tab:notations}.

\subsection{\Mbname} \label{subsec:strategy_planner}
We first introduce \name \mbname, which deduces the optimal sequence parallelism (SP) strategies and sequence assignment, maximizing training efficiency.

\subsubsection{Problem Formulation} \label{subsubsec:problem_form}
We first formulate the optimization problem of \name \mbname.
Given a data batch containing $K$ sequences $\{\mathcal{S}_k\}$ that vary in lengths, and $N$ devices with device memory budget $E$, the factors that we need to determine are: (1) the number of SP groups, (2) the parallel degree of each SP group, and (3) which SP group should each sequence be assigned to. 
Meanwhile, as sequences within different SP groups are processed concurrently, the optimization target is to minimize the maximum execution time of all SP groups.

Since the candidate set of SP degrees is very small\footnote{In common, SP degrees are set as powers of 2 to fit the ``binary structure'' of chips and networks. Besides, the highest SP degree is restricted by the number of GPUs and the context length.}, and each sequence can only be assigned to one SP group, we transform all decision variables into 0-1 integer variables. 
In particular, we assume there are $P$ virtual SP groups, where the $p^{th}$ SP group $\mathcal{G}_p$ has an SP degree of $d_p$. 
For instance, if there are 2 GPUs, we have three virtual groups with SP degrees of 1, 1, and 2, respectively.
Then, we define the group selection vector $\boldsymbol{m}=\langle m_1,m_2,...,m_P \rangle \in \{0,1\}^P$, where $m_p=1$ indicates that $\mathcal{G}_p$ is selected while $m_p=0$ means the opposite. 
By doing so, the number of SP groups and the parallel degree of each SP group can be easily described via $\boldsymbol{m}$. 
Subsequently, we further define the sequence assignment matrix $\boldsymbol{A} \in \{0,1\}^{K\times P}$, where $A_{k,p}=1$ represents that $\mathcal{S}_k$ is assigned to $\mathcal{G}_p$. 
Based on this, we can formulate a joint-optimization problem of the SP group selection and sequence assignment:
{
\begin{align}
     \quad\quad&\quad \argmin_{\boldsymbol{m} \in \{0,1\}^P; \boldsymbol{A} \in \{0,1\}^{K\times P}} \quad C 
        \label{eq:minimize_time} \\
    \text{s.t.} \quad
    & \text{Time}(\{s_k, A_{k,p}\};d_p)  \leq C, \ \forall p \in [1,P] 
        \label{cond:time} \\
    & \text{Memory}(\{s_k, A_{k,p}\};d_p) \leq E \text, \ \forall p \in [1,P] 
        \label{cond:memory} \\
    & \sum\nolimits_p{d_p\times m_p} \leq N  
        \label{cond:device_group} \\
    & \sum\nolimits_k{A_{k,p}} \leq m_p\times K, \ \forall p \in [1,P] 
        \label{cond:m_p} \\
    & \sum\nolimits_p{A_{k,p}} = 1, \ \forall k \in [1,K]
        \label{cond:assign} 
\end{align}
}

\noindent Here, $\text{Time}(\{s_k, A_{k,p}\};d_p)$ and $\text{Memory}(\{s_k, A_{k,p}\};d_p)$ denotes the execution time and the memory consumption on each device in SP group $\mathcal{G}_p$, which will be illustrated in~\S\ref{subsubsec:cost_est} in detail.
The optimization target is to minimize the maximum execution time among all SP groups (Cond.~\eqref{cond:time}).
Cond.~\eqref{cond:memory} represents the memory constraint of each device in each SP group.
Cond.~\eqref{cond:device_group} denotes that the total parallelism degrees of all selected SP groups should not be larger than the cluster device number $N$.
Cond.~\eqref{cond:m_p} ensures that no sequences will be assigned a virtual SP group that is not selected.
Cond.~\eqref{cond:assign} requires that each sequence must be assigned to one and only one group.

\begin{table}[!t]
    \centering
    \caption{\small{
        Notations used in this work.
    }}
    \label{tab:notations}
    \small
    \scalebox{.95}{
    \begin{tabular}{ll}
    \toprule
    $N$ & The number of available GPUs \\
    $P$ & The number of virtual sequence parallel (SP) groups \\
    $\mathcal{G}_p$ & The $p^{th}$ group \\
    $d_p$ & The SP degree of $\mathcal{G}_p$ \\
    $K$ & The number of sequences \\
    $\mathcal{S}_k$ & The $k^{th}$ sequence \\
    $s_k$ & The sequence length of $\mathcal{S}_k$ \\
    $Q$ & The number of buckets after sequence bucketing \\
    $\mathcal{B}_q$ & The $q^{th}$ bucket \\
    $\hat{s}_q$ & The upper limit of sequence length of $\mathcal{B}_q$ \\
    $\hat{b}_q$ & The number of sequences in $\mathcal{B}_q$ \\
    \bottomrule
    \end{tabular}
    }
\end{table}

\subsubsection{Cost Estimation} \label{subsubsec:cost_est}
To solve the optimization problem~\eqref{eq:minimize_time}, it is necessary to estimate $\text{Time}(\{s_k, A_{k,p}\};d_p)$ and $\text{Memory}(\{s_k, A_{k,p}\};d_p)$ accurately. 
Next, we then analyze the memory consumption and execution time of sequence parallelism with input sequences of variant lengths.

Memory consumption has two components: model states and forward activations.
Firstly, given a model, in Ulysses-style SP, the memory consumption of model states depends only on the ZeRO-stage applied and the number of available devices $N$. 
For instance, when ZeRO-3 is applied, the model states are evenly sharded over $N$ devices, unaffected by SP group selection or sequence assignment.
Secondly, for a device SP group $\mathcal{G}_p$ with an SP degree of $d_p$, the activation memory cost is proportional to the total number of tokens (i.e., the summed lengths of sequences) assigned to $\mathcal{G}_p$, and inversely proportional to the SP degree $d_p$. This is because sequence parallelism scatters the tokens evenly across the devices within the SP group.
Therefore, for each device in SP group $\mathcal{G}_p$, its memory consumption can be estimated as:
\begin{equation}
{
\begin{aligned}
\text{Memory}(\{s_k, A_{k,p}\};d_p) = \sum_{k} \frac{A_{k,p}s_k}{d_p} \text{M}_{token} + \text{M}_{ms},  \label{eq:memory_cost}
\end{aligned}
}
\end{equation}
\noindent where $\text{M}_{ms}$ denotes the memory consumption of model states memory that is fixed across all devices, $\text{M}_{token}$ represents the activation memory cost of each token, and $A_{k,p}$ denotes whether $\mathcal{S}_k$ is assigned to $\mathcal{G}_p$.

Previous works~\cite{DBLP:Galvatron, Galvatron-BMW, DBLP:ALPA} have proposed an effective execution cost model for distributed training of LLMs with fixed-length sequences, commonly utilizing the $\alpha$-$\beta$ model~\cite{DBLP:journals/pc/Hockney94} $T=\alpha W + \beta$ to estimate the communication and the computation overhead, where $W$ represents the workload (e.g., the computation FLOPs or communication volumes), $\alpha$ reflects the execution rate (e.g., the per-FLOP computation time or the per-byte communication time), and $\beta$ denotes the fixed overhead (e.g., kernel launch latencies).
However, existing works assume homogeneous sequence lengths and fail to accurately estimate costs for varied sequence lengths. 
Therefore, \name extends the $\alpha$-$\beta$ model, making sequence length the independent variable to handle real-world training corpora.
Specifically, to adapt to Transformer models, we model the computation cost of the attention mechanism and the other modules separately. 
The reason is that the computation cost of attention mechanism is positively correlated with the quadratic of sequence length, while the other modules like linear projection have a linear computation cost w.r.t. the sequence length. 
Besides, sequence parallelism scatters the computation across the devices within an SP group $\mathcal{G}_p$, thus the per-device computation volume is inversely proportional to SP degree $d_p$. 
Therefore, by summing the computation cost of all assigned sequences, we estimate the computation overhead as follows:
\begin{equation}
{
\begin{aligned}
\text{T}_{comp}(\{s_k, A_{k,p}\};d_p) = \frac{1}{d_p} \sum_{k} A_{k,p} (\alpha_1 s_k^2+\alpha_2 s_k) + \beta_1,
    \label{eq:comp}
\end{aligned}
}
\end{equation}
where $\alpha_1, \alpha_2, \beta_1$ denote the coefficients of the $\alpha$-$\beta$ model for computation cost, which are obtained through profiling. 

The communication volume of Ulysses-style SP mainly comes from the All-to-All communication, whose volume is proportional to the sequence length $s_k$ and inversely proportional to SP degree $d_p$~\cite{DBLP:deepspeed-sp-Ulysses}.
Hence, \name estimate the All-to-All communication cost as follows:
\begin{equation}
{
\begin{aligned}
\text{T}_{comm}(\{s_k, A_{k,p}\};d_p) = \frac{1}{d_p v_p} \sum_{k} A_{k,p} \alpha_3 s_k + \beta_2,
    \label{eq:comm}
\end{aligned}
}
\end{equation}
where $\alpha_3, \beta_2$ are coefficients given by profiling, and $v_p$ represents the interconnect bandwidth of the devices within $\mathcal{G}_p$, which can also be profiled out.

As can be seen, \name draws inspiration from the $\alpha$-$\beta$ model, using $\beta_\cdot$ to represent the data-independent startup latency, while utilizing the $\alpha_\cdot$ to fit the time cost for both communication and computation process according to their respective behaviors. 
Then, we combine them to estimate the overall execution time of sequence parallelism with varied-length sequences as follows:
\begin{equation}
{
\begin{aligned}
\text{Time}(\{s_k, A_{k,p}\};d_p) = \text{T}_{comp} + \text{T}_{comm}.
    \label{eq:time_cost}
\end{aligned}
}
\end{equation}

Furthermore, when combining sequence parallelism with ZeRO (especially ZeRO-3), we also estimate the overhead of parameter gathering and gradient synchronization, and also consider the overlapping of computation and communication like previous works~\cite{DBLP:Galvatron, Galvatron-BMW}. 
As ZeRO is orthogonal to our work, and its overhead is unrelated with the sequence parallelism nor the sequence lengths, we omit such details in this paper for clarity.
Experiments show that the overall cost estimation error is below 6\%, detailed in \ifasplos\else{Appendix~\ref{sec:appendix_est_acc}.}\fi


\subsubsection{Problem Solving}
According to the problem formulation in~\S\ref{subsubsec:problem_form} and the overhead estimation in~\S\ref{subsubsec:cost_est}, we can find that all the constraints and the optimization target is linear with respect to the decision variables $m_p$ and $A_{k,p}$.
Therefore, the optimization problem~\eqref{eq:minimize_time} turns out a Mixed-Integer Linear Programming (MILP) problem. 
Although existing advanced MILP solvers like SCIP~\cite{BolusaniEtal2024OO} are capable of solving MILP problems, 
the number of decision variables in problem~\eqref{eq:minimize_time} is too large and uncontrollable, making it too complex to derive feasible solutions within a reasonable time. 
To tackle this obstacle, we need to simplify the problem to decrease the number of decision variables.

In particular, since the number of decision variables is proportional to the number of sequences, and sequences with similar lengths should incur similar overhead, we opt to group the sequences into a small number of buckets. 
In other words, given the sequences with various lengths, we group the sequences with similar lengths in the same bucket and represented by a unified sequence length (typically, the maximum sequence length within the bucket). 
Although this will introduce certain estimation biases, it can significantly reduce the number of unique sequence lengths and thereby lower the problem complexity. 
Below, we introduce our sequence bucketing algorithm.


A na\"ive method for sequence bucketing is to set a fixed length interval for each bucket and use its upper limit to represent the length of sequences within the bucket. 
For instance, the upper limit of sequence length can be set as multiples of 2K, that is, 0-2K, 2K-4K, 4K-6K, and so on, forming several buckets. 
However, as discussed in \S\ref{sec:obs}, the sequence lengths in real-world datasets exhibit a complex long-tail distribution rather than a uniform distribution. Besides, different datasets exhibit distinct distributions. Consequently, such a na\"ive bucketing method would inevitably introduce large estimation biases and cannot be generalized. 

To reduce the estimation biases caused by bucketing, we adopt an adaptive sequence bucketing mechanism and propose a dynamic programming algorithm to minimize the bucketing deviation.
Specifically, given $K$ sequences $\{\mathcal{S}_k\}$, we group them into $Q$ buckets, where the $q^{th}$ bucket $\mathcal{B}_q$ has the upper limit of sequence length $\hat{s}_q$, and containing all sequences satisfying $\hat{s}_{q-1} < s_k \leq \hat{s}_q$.
The bucketing error can be measured as the total deviation of the sequence length to the upper limit of the bucket it belongs to, and the optimization target of sequence bucketing can be defined as:
\begin{equation}
    \argmin_{\{\hat{s}_q\}} 
    \sum_q \sum_k I[\hat{s}_{q-1} < s_k \leq \hat{s}_q] (\hat{s}_q - s_k).
\end{equation}
We solve this bucketing problem via a dynamic programming algorithm.
We first sort sequences in ascending order of sequence lengths, i.e., $s_1 \leq s_2 \leq ... \leq s_K$, and then define $err[k][q]$ as the minimized error of bucketing the first $k$ sequences into $q$ buckets. Then, starting with $err[0][0]=0$, we can derive the following state transition formula of dynamic programming:
\begin{equation} \label{eq:seq_bkt_dp}
    err[k][q]=\min_{j \in [0, k-1]}{\{err[j][q-1]+\sum_{i=j+1}^k(s_k-s_i)}\}.
\end{equation}
Here, $\sum_{i=j+1}^k(s_k-s_i)$ denotes the bucketing error of the $q^{th}$ bucket when selecting $\hat{s}_{q-1} = s_j$ as the upper limit of the ${(q-1)}^{th}$ bucket $\mathcal{B}_{q-1}$. 
Through this dynamic programming algorithm, we determine the bucket boundaries that minimizes the bucketing error adaptively to the data, and group the sequences $\{\mathcal{S}_k\}$ into $Q$ buckets $\{\mathcal{B}_q=\{\mathcal{S}_{k_q}\}\}$.
In practice, we set bucket number $Q$ as 16 by default.


We now re-formulate the optimization problem based on the bucketed sequences.
Given the number of available GPUs $N$, the device memory capacity $E$, and $K$ sequences $\{\mathcal{S}_k\}$ as well as $Q$ sequence buckets $\{\mathcal{B}_q=\{\mathcal{S}_{k_q}\}\}$, where bucket $\mathcal{B}_q$ has $\hat{b}_q$ sequences and upper length limit $\hat{s}_q$, we keep the definition of SP groups as problem~\eqref{eq:minimize_time}, and re-define the sequence assignment matrix $\boldsymbol{\hat{A}} \in \mathbb{N}_{\geq 0}^{Q \times P}$ such that $\hat{A}_{q,p}$ represents the number of the sequences in the $q^{th}$ bucket $\mathcal{B}_q$ assigned to the $p^{th}$ SP group $\mathcal{G}_p$.
Then, we can re-formulate the optimization problem as follows:
{
\begin{align}
    & \quad\quad\quad \argmin_{\boldsymbol{m} \in {\{0, 1\}}^P; \boldsymbol{\hat{A}} \in \mathbb{N}_{\geq 0}^{Q \times P}} \quad C 
        \label{eq:minimize_time_bkt} \\
    \text{s.t.} \quad
    & \text{Time}(\{\hat{s}_q, \hat{A}_{q,p}\};d_p)  \leq C, \ \forall p \in [1,P] 
        \label{cond:time_bkt} \\
    & \text{Memory}(\{\hat{s}_q, A_{q,p}\};d_p) \leq E, \ \forall p \in [1,P] 
        \label{cond:memory_bkt} \\
    & \sum\nolimits_p{d_p \times m_p} \leq N  
        \label{cond:device_group_bkt} \\
    & \sum\nolimits_q{\hat{A}_{q,p}} \leq m_p \times K, \ \forall p \in [1,P] 
        \label{cond:m_p_bkt} \\
    & \sum\nolimits_p{\hat{A}_{q,p}} = \hat{b}_q, \ \forall q \in [1,Q]
        \label{cond:assign_bkt} 
\end{align}
}
where Cond.~\eqref{cond:assign_bkt} ensures that all the sequences in bucket $B_q$ are assigned. 
It is obvious that the re-formulated optimization problem~\eqref{eq:minimize_time_bkt} is also a MILP problem. 
In practice, \name utilizes SCIP, an advanced MILP solver library, to solve the problem~\eqref{eq:minimize_time_bkt}. 
After obtaining the optimal group selection vector $\boldsymbol{m}^*$ and the optimal sequence assignment matrix $\hat{\boldsymbol{A}}^*$, we can derive the optimal parallelism plan according to $\boldsymbol{m}^*$ and dispatch the training sequences across the SP groups according to $\hat{\boldsymbol{A}}^*$. 
The solving time of problem~\eqref{eq:minimize_time_bkt} is typically within 5-15 seconds, which can be overlapped with the training time of one batch (\S\ref{sec:imple}).

\subsection{\Gbname} \label{subsec:seq_chunk}
When the input batch contains too many sequences, they cannot be processed together due to the limited memory capacity, and therefore the optimization problem~\eqref{eq:minimize_time_bkt} will have no feasible solutions due to memory constraint~\eqref{cond:memory_bkt}.
Gradient accumulation is the common technique for such cases, which splits the global data batch into several micro-batches, executes each micro-batch sequentially and accumulates the model gradients for parameter update. 
For training systems intended for homogeneous sequence lengths, micro-batch chunking is straightforward --- we can simply fix the number of sequences in each micro-batch.  
However, in our scenario where input sequences are associated with heterogeneous lengths, micro-batch chunking is non-trivial. 
Therefore, \name designs a \gbname to blast the sequences into micro-batches for \mbname to determine the optimal sequence parallelism strategies.

Given input data batch $\mathcal{B}=\{\mathcal{S}_k\}$ with $K$ sequences, the \gbname blasts the sequences into $M$ disjoint micro-batches $\{\mathcal{M}_i\}$, satisfying $\bigcup_{i=1}^M \mathcal{M}_i = \mathcal{B}$.
In the following, we summarize several propositions based on theoretical analysis and empirical observations, and introduce our designs of \gbname motivated by these propositions.

\underline{\textit{Takeaway \#1:}} 
\textit{In most cases, having fewer micro-batches is likely to be more efficient.}

This takeaway can be deduced from the cost estimation in~\S\ref{subsubsec:cost_est}.
Either in computation or communication, there is a fixed overhead term denoted as $\beta$ that exists for each micro-batch execution, so having more micro-batches introduces more additional overhead. 
Besides, if we have many micro-batches, which implies that each micro-batch only consists of very few tokens, then the workload distributed to each micro-batch may not be sufficient to fully utilize either the computation capacity or the communication bandwidth. 
Therefore, a smaller number of micro-batch number $M$ usually gives better efficiency.

However, this does not mean that the smallest $M$ always achieves the best performance.
Hence, our \gbname first calculates the smallest feasible micro-batch number $M_{min} = \left\lceil\frac{Batch\_Total\_Token}{Cluster\_Token\_Capacity}\right\rceil$.
Then, it traverses the micro-batch number in range $[M_{min}, M_{min} + M')$ to find the best one, where $M'$ is the number of trails (5 by default).

\underline{\textit{Takeaway \#2:}} 
\textit{A smaller variance of sequence lengths within a micro-batch is likely to be more efficient.}

This takeaway is based on both the theoretical analysis of execution overhead (\S\ref{subsubsec:cost_est}) and empirical observations derived by solving the optimization problem~\eqref{eq:minimize_time_bkt}.
Specifically, the memory consumption (Eq.~\eqref{eq:memory_cost}) is linear to the sequence length, while the computation time (Eq.~\eqref{eq:comp}) is quadratic to sequence length.
Consequently, as the sequence length $s_k$ increases, the computation overhead increases faster than the memory consumption, leading to imbalance between computation and memory cost.
For instance, for two sequences $\mathcal{S}_1$, $\mathcal{S}_2$ with length $s_1=4K$, $s_2=16K$ within one micro-batch, $\mathcal{S}_1$ is assigned to SP group $\mathcal{G}_1$ with $d_p=8$, while $\mathcal{S}_2$ is assigned to $\mathcal{G}_2$ with $d_p=32$.
Although the memory consumption of $\mathcal{G}_1, \mathcal{G}_2$ is the same, the computation cost of $\mathcal{G}_2$ is larger than that of $\mathcal{G}_1$ due to the quadratic computation volume of long sequence $\mathcal{S}_2$, which requires $\mathcal{G}_1$ to wait for $\mathcal{G}_2$ to finish and causes computation resource wastage.
On the other hand, if we try to align the computation time of sequences with diverse lengths, their memory consumption will be distinct, leading to memory under-utilization.
To conclude, larger variance of sequence lengths within a micro-batch leads to resource under-utilization of either computation or memory.

Motivated by this, \name \gbname first sorts the input sequences according to their lengths, and ensures that sequences with smaller variance of lengths are blasted into one micro-batch.

\underline{\textit{Takeaway \#3:}} 
\textit{The total token number of each micro-batch should be made as evenly distributed as possible.}

This takeaway focuses on striking a balance in memory consumption across micro-batches, which is proportional to the total token number. 
It is designed to prevent potential out-of-memory (OOM) situations when splitting micro-batches and also to avoid under-utilization of device memory.
This guidance also contributes to takeaway \#1, as imbalanced token blasting leads to more micro-batches.
Therefore, we design a memory-balanced micro-batch chunking algorithm based on dynamic programming, detailed in \ifasplos\else{Appendix~\ref{sec:appendix_mb_chunk}.}\fi


\subsection{Overall Workflow of \name Solver} \label{subsec:solver_workflow}
We now introduce the overall workflow of \name solver, as illustrated in Alg.~\ref{alg:solver}.
Given the data batch $\mathcal{B}$,
we first calculate the minimum feasible micro-batch number $M_{min}$ in Line~\ref{line:min_mb} based on cluster memory capacity, as discussed in~\S\ref{subsec:seq_chunk}, and then traverse micro-batch number $M$ starting from $M_{min}$ in Line~\ref{line:loop_mb_num}.
For each traversed $M$, the \gbname (\S\ref{subsec:seq_chunk}) is invoked to blast the sequences into $M$ micro-batches $\{\mathcal{M}_i\}$, $i \in [1,M]$ (Line~\ref{line:seq_blaster}).
Subsequently, for each micro-batch $\mathcal{M}_i$, we first group sequences into $Q$ buckets (Line~\ref{line:seq_bucketing}) and then utilize the \mbname (Line~\ref{line:parallel_planner}) to optimize the sequence parallelism strategies for the current micro-batch data, which solves the MILP problem as discussed in~\S\ref{subsec:strategy_planner}.
Line~\ref{line:p_acc} gathers the optimal time and strategy of each micro-batch to form the results for the whole data batch $\mathcal{B}$, and Line~\ref{line:opt_plan} finds the best parallelism plan $\mathcal{P}^*$ with minimum execution time $T^*$ across various tries of micro-batch number.

To improve the efficiency of the solver, \name employs a two-level multi-process solving technique to parallelize the solving process.
Specifically, \name explores various micro-batch numbers in parallel (Line ~\ref{line:loop_mb_num}) with multiple processes, and optimizes the parallelism strategies of each micro-batch in parallel (Line~\ref{line:loop_microbatch}) as well.
Through this technique, the solving overhead of the \name solver is close to the overhead of solving one round of the MILP problem in~\S\ref{subsec:strategy_planner}, typically within 5-15 seconds, and is independent of the number of sequences nor the number of micro-batches.
Therefore, the solving efficiency of \name solver is guaranteed.

\begin{algorithm}[t]
\caption{\name Solver Workflow}
\label{alg:solver}
    \scalebox{0.95}{
    \begin{minipage}{\textwidth}
\LinesNumbered
\KwIn{Data Batch $\mathcal{B}=\{\mathcal{S}_k\}$ with $K$ sequences, \# Buckets $Q$, \\
\# Devices $N$, Device Memory Capacity $E$, \# Trails $M'$}
\KwOut{Minimized time $T^*$, Parallelism Plan $\mathcal{P}^*$}
$T^*, \mathcal{P}^* \gets \infty, None$\;
$M_{min} \gets \text{get\_min\_microbatch\_num}(\mathcal{B}, N, E)$\; \label{line:min_mb}
\For{$M$ in $M_{min}, M_{min}+1,\ldots,M_{min}+M'-1$ 
in parallel
}{  \label{line:loop_mb_num}

$T_{\mathcal{B}}, \mathcal{P}_{\mathcal{B}} \gets 0, []$\;
$\{\mathcal{M}_i\} \gets \text{Sequence\_Blaster}(\mathcal{B}, M)$\; \label{line:seq_blaster}
\For{$\mathcal{M}$ in $\mathcal{M}_1,\mathcal{M}_2,\ldots,\mathcal{M}_M$ 
in parallel
}{ \label{line:loop_microbatch}

$\{B_q\} \gets \text{Sequence\_Bucketing}(\mathcal{M},Q)$\; \label{line:seq_bucketing}
$T_{\mathcal{M}}, \mathcal{P}_{\mathcal{M}} \gets \text{Parallelism\_Planner}(\{B_q\},N,E)$\; \label{line:parallel_planner}
$T_{\mathcal{B}} \gets T_{\mathcal{B}} + T_{\mathcal{M}};$ $\mathcal{P}_{\mathcal{B}}.extend(\mathcal{P}_{\mathcal{M}})$\; \label{line:p_acc}
}
\If{$T_{\mathcal{B}} < T^*$}{ 
$T^*, \mathcal{P}^* \gets T_{\mathcal{B}}, \mathcal{P}_{\mathcal{B}}$\; \label{line:opt_plan}
}
}
\Return $T^*, \mathcal{P}^*$\;
    \end{minipage}
    }
\end{algorithm}

\begin{figure*}[t]
    \centering
    \includegraphics[width=1.\linewidth]{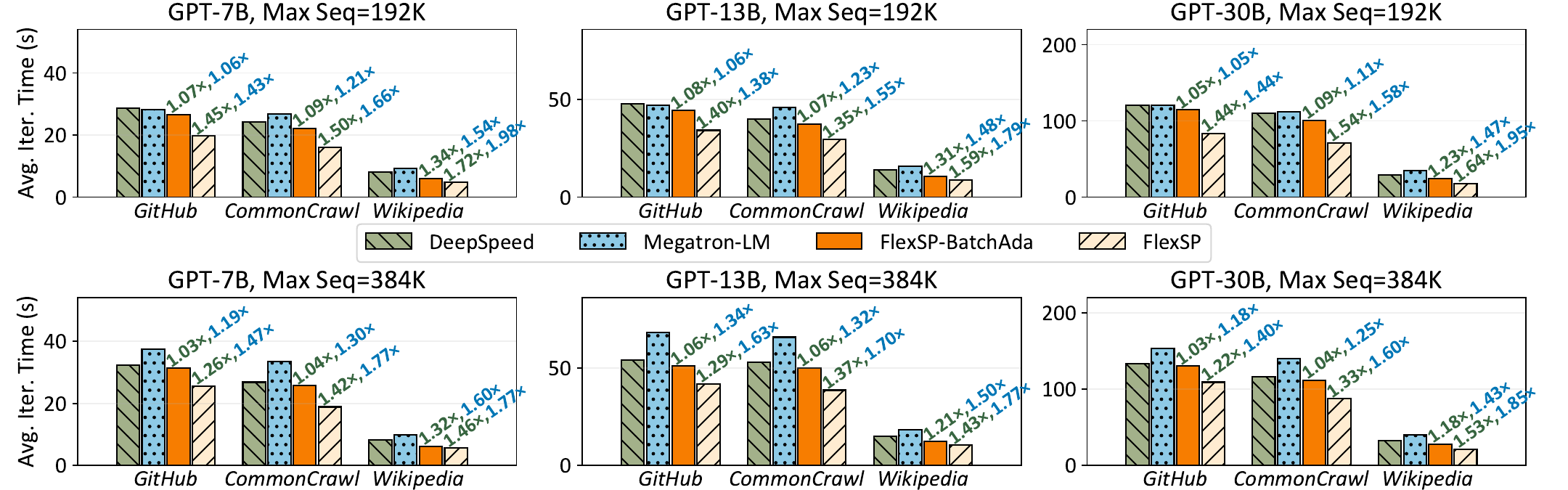}
    \caption{
        End-to-end evaluation (in seconds per iteration) for specific model sizes and maximum context lengths (Max Seq) across three datasets, shown in each sub-figure.
        Speedup ratios compared to DeepSpeed (green, left) and Megatron-LM (blue, right) are indicated.
    }
    \label{fig:main-result}
\end{figure*}

\section{Implementation} \label{sec:imple}
We build the proposed method as an efficient LLM training system, \name.
We implement the \name solver with Python and C++, leveraging the SCIP~\cite{BolusaniEtal2024OO} library for solving the MILP problem. 
We then develop the \name runtime engine on top of PyTorch for executing the training process based on the strategies optimized by the solver. 
In our implementation, we use NCCL~\cite{nccl} as the communication backend.
As for the attention kernel, we utilize the state-of-the-art flash-attn~\cite{DBLP:flash_attn, DBLP:flash_attn_2} library's interface for varied-length sequence packing to perform attention computation.
As for parallelisms, we implement the Ulysses-style SP similar to DeepSpeed-Ulysses~\cite{DBLP:deepspeed-sp-Ulysses} and implement ZeRO with PyTorch FSDP~\cite{DBLP:pytorch-fsdp}.
We now introduce several key points in our implementation of \name for efficient training.

\paragraph{Hot Switching and Group Management}
\name implements sequence parallelism in a hot switching manner to deal with the varied parallelism strategies for the distinct input data.
Given each micro-batch as well as the corresponding parallelism strategies, \name generates the SP communication groups dynamically and scatters the data into the corresponding group.
In order to avoid redundant creation and storage overhead of communication groups, \name maintains a NCCL group pool to manage the complicated SP groups.
Specifically, \name generates communication groups on the fly, and new groups are created only when necessary, while existing ones are reused to optimize resource usage.
Therefore, in FlexSP, dynamically adjusting the SP groups does not incur any overhead if the groups are cached. 
The number of communication groups needed for each GPU is up to $\log N$, where $N$ is the number of GPUs.\footnote{Because the sizes of SP groups are always powers of 2, we let each GPU to always pair with its neighbors. 
For example, with $N=4$ GPUs, there are at most 3 communication groups, i.e., [0,1], [2,3], and [0,1,2,3], with each GPU associated with 2 ($=\log N$) groups.}
In our evaluation, creating $\log64=6$ communication groups takes under 10 seconds, negligible compared to the overall training time.

\paragraph{Disaggregating Solving and Training}
For each training data batch, there are two phases, i.e., the solver deduces the optimal plan (i.e., a combination of parallelism strategies and sequence assignment) and the executor carries out the training. 
Since the problem solving is on CPUs while the training is on GPUs, we disaggregate the two phases to facilitate overlapping. 
In particular, on each GPU node (machine), we establish a service of the solver, which takes as input the lengths of one data batch, and runs Alg.~\ref{alg:solver} to deduce the optimal plan. 
Subsequently, we manage a distributed storage to gather the plans, and the executor sequentially reads one plan per iteration to train.
By doing so, \name solves the problem for multiple data batches concurrently, and the problem solving is also overlapped with the training process.


\section{Experiments} \label{sec:experiments}

\subsection{Experimental Setups}

\subsubsection*{Baseline Systems}
We compare our system with the state-of-the-art (SOTA) distributed LLM training systems, i.e., Megatron-LM~\cite{DBLP:megatron} and DeepSpeed~\cite{Deepspeed}.
Megatron-LM supports 4D-parallelism, including TP (with Megatron-style SP), PP, DP (ZeRO-1), and CP.
DeepSpeed supports DeepSpeed-ZeRO and Ulysses-style SP.
Both these systems are tailored for homogeneous sequence lengths and only support training LLMs with a single, static parallelism strategy.
Our system, namely \name, is adaptive to the heterogeneous workloads of varied-length sequences, and is capable to generate the optimal sequence parallelism strategies adaptively.

For further evaluation of our adaptive feature, we also introduce a variant of \name as a baseline, \name-BatchAda.
Unlike DeepSpeed which employs one static strategy along the whole training process, \name-BatchAda adaptively applies the most efficient homogeneous SP strategy for each data batch, e.g., two SP=32 groups for the first batch and eight SP=8 groups for the second batch.
Compared to \name-BatchAda, \name not only allows adaptive strategies across data batches, but also supports heterogeneous SP strategies within each data batch, e.g., mixing and executing one SP=32 group and four SP=8 groups concurrently.

\subsubsection*{Hardware Environments}
We conduct all the experiments on a GPU cluster with 8 nodes, with each node consisting of 8 NVIDIA A100 40GB GPUs equipped with NVLink. 
All nodes are interconnected by 400Gbps InfiniBand network.

\subsubsection*{Experimental Workloads}
We conduct experiments on GPT-series LLMs of three different sizes, GPT-7B, GPT-13B, GPT-30B.
Refer to \ifasplos\else{Appendix~\ref{sec:appendix_model_config} }\fi for more details.
We choose three different datasets, including \textit{GitHub}, \textit{CommonCrawl}, and \textit{Wikipedia}.
Fig.~\ref{fig:seq-dist} displays the distribution of sequence lengths of these datasets.
We also evaluate each system on these datasets under different maximum context length limits, i.e., 384K and 192K. 
The sequences exceed the maximum context length limit will be eliminated during training.

\subsubsection*{Protocols}
We apply sequence packing for all systems.
Specifically, for baseline systems Megatron-LM, DeepSpeed and \name-BatchAda, we use the \textit{Best-fit Packing}~\cite{icml-bestfitpack} as introduced in \S\ref{sec:preliminary}. 
For \name, the solver will automatically determine the sequence packing.
As for the parallelism strategy, we manually tune the most efficient strategy for baseline systems under different workloads, including parallelism degrees of DP, TP, PP, CP, SP.
We also apply activation checkpointing strategies for each system to accommodate model training with a context length of 384K.
We fix the global batch size of each training step as 512 for all workloads, and record the average iteration time over 40 iterations after 10-iteration's warm-up.
Refer to \ifasplos\else{Appendix~\ref{sec:appendix_parallel_config} }\fi for details.

\subsection{End-to-End Performance}
We compare the end-to-end performance of each system in Fig.~\ref{fig:main-result}, which shows the average iteration time of each system across different workloads. 
The results demonstrate that across all the model sizes, datasets, and context lengths, \name consistently outperforms all baseline systems, achieving a maximum speedup of 1.72$\times$ compared to DeepSpeed and 1.98$\times$ compared to Megatron-LM.

We first analyze the performance gain of \name compared to SOTA systems.
The advantages of \name primarily arise from the communication gains achieved by its flexible sequence parallelism strategy.
As mentioned in \S\ref{sec:obs}, the parallelism group needs to be large enough to shard excessively long sequences to fit the model into device memory.
For instance, under 384K maximum context length, DeepSpeed requires SP=64 while Megatron requires TP=16, CP=4 or TP=8, CP=8.
Such large parallelism groups must communicate with slow inter-node network bandwidth, thus leading to inefficient communication.
SOTA systems maintain a homogeneous and static parallelism strategy along the training process, forcing all sequences in the dataset to utilize the large groups with slow inter-node bandwidth, which is inefficient for shorter sequences.
On the contrary, \name allows shorter sequences to enjoy the higher communication efficiency within smaller parallelism groups, while maintaining larger groups for long sequences to satisfy the memory constraint.
For instance, \name may assign a sequence with 100K into a group with SP=32 to avoid OOM errors, while scattering sequences with 16K into a group with SP=8 to enjoy the fast intra-node connection.
Such flexible strategy effectively reduces the communication overhead and contributes to the system efficiency of \name.

\iftrue
\begin{table}[!t]
    \centering
    \captionof{table}{
        Details of heterogeneous SP groups employed in each micro-batch of each case.
        Each $d \times m$ indicates we form $m$ SP=$d$ groups, and each $\langle \cdots \rangle \times x$ indicates the set of heterogeneous SP groups is employed for $x$ micro-batches ($\times 1$ is omitted). 
    }
    \label{tab:case-study}
    \scalebox{.975}{
    \begin{tabular}{c|c|c}
    \toprule
              & Case 1           & Case 2           \\ \hline\hline
    DeepSpeed & $\langle 64\rangle\times 5$ & $\langle 64\rangle\times 7$ \\ \hline
    \name-BatchAda  & $\langle 16\times 4\rangle\times 5$ & $\langle 32\times 2\rangle\times 7$ \\ \hline
    \name     & \makecell{$\langle 32,16,8\times 2\rangle$\\$\langle 8\times 8\rangle\times 2$\\$\langle 8\times 7,4\times 2\rangle$\\$\langle 1\times 64\rangle$} 
              & \makecell{$\langle 64\rangle$\\$\langle 32,16\times 2\rangle$\\$\langle 16\times 3,8\times 2\rangle$\\$\langle 8\times 8\rangle\times 2$\\$\langle 1\times 64\rangle$} \\
    \bottomrule
    \end{tabular}
    }
\end{table}
\fi

Furthermore, the strength of \name is correlated with the long-tail distribution of sequence lengths --- a more pronounced long-tail leads to greater communication benefits, resulting in a more significant speedup.
As shown in Fig.~\ref{fig:seq-dist}, the \textit{Wikipedia} dataset has the greatest skewness in three datasets. 
Over 96\% of the sequences in \textit{Wikipedia} are below 8K, considerably greater than those in \textit{GitHub} and \textit{CommonCrawl}, and the proportion of sequences exceeding 32K is much smaller than those in the other datasets.
Compared to SOTA systems, such great skewness benefits \name to achieve speedup of up to 1.98$\times$ on \textit{Wikipedia}, while the speedups on \textit{CommonCrawl} and \textit{GitHub} are slightly lower, up to 1.77$\times$ and 1.63$\times$, respectively.

Then, we analyze the performance of \name-BatchAda, which employs homogeneous strategy within each data batch but allows adaptive strategies across data batches.
It also gains benefits from communication and achieves speedup ratio up to 1.34$\times$ and 1.60$\times$ compared to DeepSpeed and Megatron-LM, respectively.
However, due to its homogeneous strategy within each batch, its performance gain on \textit{GitHub} and \textit{CommonCrawl} is relatively low, as these datasets possess long sequences in many data batches, forcing these batches to use large and inefficient parallelism groups.
In comparison, \name allows heterogeneous and adaptive strategies at a nuanced granularity, both among and within data batches, further increasing the potential for reducing communication overhead and achieving acceleration up to 1.42$\times$ compared to \name-BatchAda.

\iftrue
\begin{figure}[!t]
    \begin{minipage}{.57\linewidth}
        \flushleft
        \includegraphics[width=.95\linewidth]{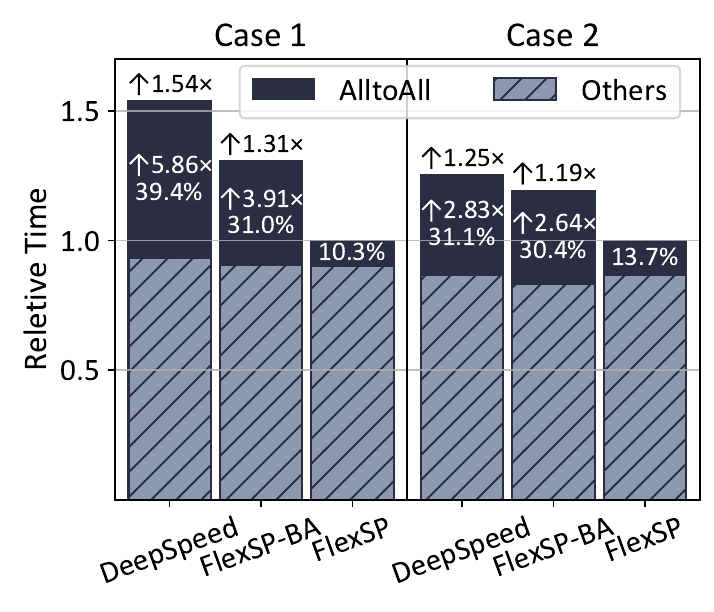}
        \vspace{-5pt}
        \subcaption{
        \small{
            }
        }
        \label{fig:case-study}
    \end{minipage}    
    \hfill
    \begin{minipage}{.42\linewidth}
        \centering
        \includegraphics[width=\linewidth]{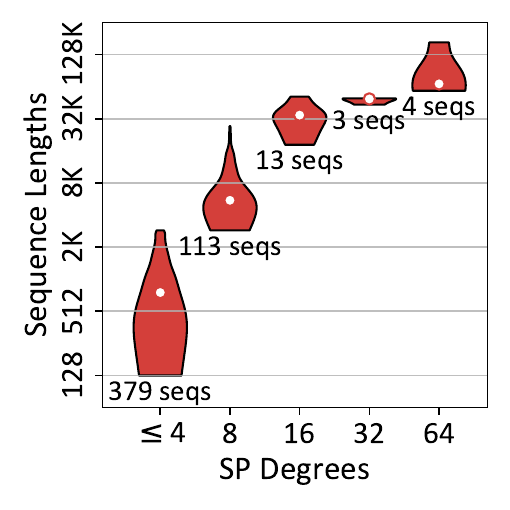}
        \vspace{-8pt}
        \subcaption{
        \small{
            }
        }
        \label{fig:case-study-seq-assign}
    \end{minipage}
    \vspace{-5pt}
    \caption{
    (\ref{fig:case-study}) Breakdown of end-to-end time (All-to-All+Others) in case study (BA is short for BatchAda).
    (\ref{fig:case-study-seq-assign}) Distribution of sequence lengths assigned to different SP degrees in Case 2, visualized as a violin plot. The white circle indicates the median.
    }
\end{figure}

\fi

\subsection{Case Study}
To analyze the performance gains of \name more clearly, we conduct an in-depth case study on two iterations of GPT-7B on \textit{CommonCrawl} with a maximum context length of 384K.

We present the parallelism strategies, i.e., details of the employed SP groups, in Tab~\ref{tab:case-study}.
We also break down the end-to-end time and highlight the portion of All-to-All communication in Fig~\ref{fig:case-study}. 
It can be seen that the major difference lies in the All-to-All communication overhead, which is the source of \name's performance gain. 
For DeepSpeed, its All-to-All communication accounts for up to 40\% of the total runtime, which is due to the large SP group (SP=64) and the limited inter-node bandwidth across 8 nodes.
\name-BatchAda adapts the SP strategies for batches (four SP=16 groups for Case 1 and two SP=32 groups for Case 2), and reduces communication cost compared to DeepSpeed, especially in Case 1 where SP=16 cuts down the communication to 31\%.
\name further optimizes communication through adaptive strategies at a finer granularity, leveraging smaller SP groups (e.g., SP=1, 4, 8) to process shorter sequences,
which significantly reduces communication over low-bandwidth inter-node connections, cuts down the All-to-All time to around 10\%, and achieves a reduction of up to 5.86$\times$ in All-to-All time and a 1.54$\times$ speedup in overall end-to-end time.

To further explore \name's flexible strategy, we present the distribution of sequence lengths assigned to different SP degrees in Case 2, as shown in Fig.~\ref{fig:case-study-seq-assign}.
In \name, sequences of diverse lengths are assigned to appropriate SP degree groups, with shorter sequences showing a clear preference to lower SP degrees so that the All-to-All communication cost can be minimized. 
Meanwhile, due to the long-tail property of the datasets, (relatively) short sequences may be routed to SP groups with (relatively) higher parallel degree, striking a good balance across all SP groups. 
This highlights the effectiveness of \name solver in optimizing the flexible parallelism strategies for sequences with varied lengths.

\begin{figure}[!t]
    \includegraphics[width=.97\linewidth]{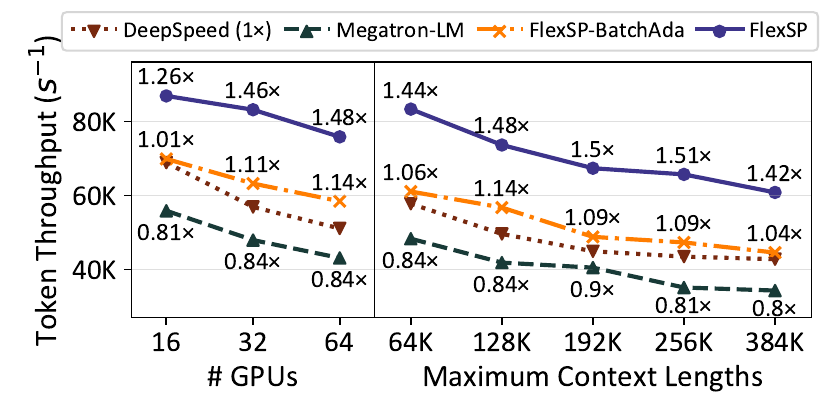}
    \caption{
    Scalability study measured as token throughput per GPU. The speedup rates are measured w.r.t. DeepSpeed.
    }
    \label{fig:scale-analysis}
\end{figure}

\subsection{Scalability Study}
To evaluate the scalability of each system, we conduct experiments on \textit{CommonCrawl}, varying both cluster size and maximum context length.
The results, measured as token throughput per GPU, are presented in Fig.~\ref{fig:scale-analysis}.

\subsubsection*{Scalability w.r.t. \# GPUs.}
We begin by evaluating performance across GPU clusters with 16, 32, and 64 GPUs, with a maximum context length of 128K.
The results indicate that \name consistently outperforms other systems, achieving a maximum speedup of 1.48$\times$ compared to DeepSpeed.
Furthermore, we find that as the cluster size increases, the reduced inter-node bandwidth brings negative impact on training throughput.
For instance, when scaling from 16 to 32 GPUs, DeepSpeed and Megatron-LM only achieve sublinear speedup of 1.65$\times$ and 1.71$\times$, respectively.
This is because the bandwidths of 32 and 64 GPUs on our cluster are lower than that of 16 GPUs, which leads to poor scalability of SOTA systems.
However, \name is much more robust to such bandwidth decrease, achieving 1.91$\times$ when scaling from 16 to 32 GPUs and 1.82$\times$ from 32 to 64 GPUs.
\name's sound scalability attributes to the adaptive strategies and the utilization of the high bandwidth of intra-node NVLink.

\subsubsection*{Scalability w.r.t. maximum context length.}
We extend the evaluation on 64 GPUs with maximum context lengths ranging from 64K to 384K.
The token throughput of all systems tends to decrease due to the increased computational FLOPs associated with longer sequences.
\name consistently maintains its optimal performance under different context length limit, achieving a speedup ratio between 1.42$\times$ and 1.51$\times$.
Furthermore, we find the speedup ratio for 64K and 384K is slightly lower than that for 256K, which is reasonable. 
For a shorter context length limit, such as 64K, the long-tail property of the dataset is weakened, resulting in fewer opportunities for adaptive optimization. 
On the other hand, for a longer context length, like 384K, the computation overhead of extremely long sequences consumes a significant amount of time, which also reduces the speedup.

\begin{table}[t]
\centering
\caption{Token estimation bias of bucketing methods.}
\begin{tabular}{l|c|c|c}
\hline
Token Error & \textit{Github} & \textit{CommonCrawl} & \textit{Wikipedia} \\
\hline
DP Bucketing & \textbf{0.7\%} & \textbf{0.5\%} & \textbf{2.3\%} \\
Naïve Bucketing & 13.4\% & 8.8\% & 22.1\% \\
\hline
\end{tabular}
\label{tab:bkt_estimation_error}
\end{table}

\begin{figure}[!t]
    \centering
    \includegraphics[width=1.\linewidth]{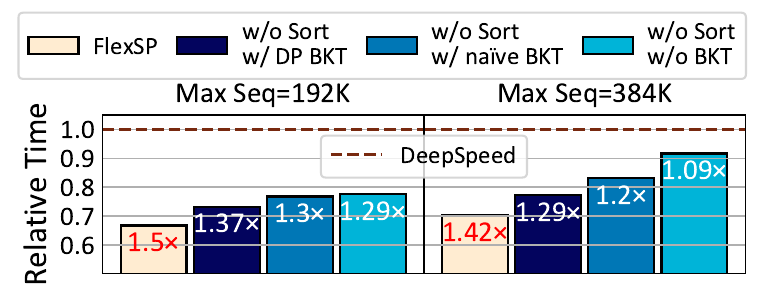}
    \caption{
        Ablation studies.
        \name adopts sequence sorting (Sort) in sequence blaster and DP bucketing (BKT) algorithm.
    }
    \label{fig:ablation}
\end{figure}

\subsection{Ablation Study}

To evaluate the efficacy of key components within the \name solver, i.e., the dynamic programming (DP) sequence bucketing in \mbname (\S\ref{subsec:strategy_planner}), and the sequence sorting mechanism in \gbname (\S\ref{subsec:seq_chunk}), we compare the performance of complete version of \name against various ablated versions on \textit{CommonCrawl}, as shown in Fig.~\ref{fig:ablation}.
Sequence sorting in sequence blaster helps reduce sequence length variance within each micro-batch.
Disabling this mechanism negatively impacts overall performance.
Additionally, replacing the DP sequence bucketing with a na\"ive even-sized bucketing introduces more biases into the bucketing estimation, leading to worse performance.
Finally, removing the bucketing mechanism entirely increases the complexity of the MILP problem, causing the solver to fail in producing a satisfactory solution within limited time.

We also evaluate the token estimation bias of the naïve bucketing and our dynamic-programming-based (DP) optimal sequence bucketing in \mbname (\S\ref{subsec:strategy_planner}). 
We show the maximum token error ratio, i.e., error token number divided by total token number, on different datasets in Tab.~\ref{tab:bkt_estimation_error}. 
We find that our optimal bucketing algorithm effectively reduces the estimation error to lower than 2.3\%, while naïve bucketing introduces error up to 22\%.
\subsection{Complexity and Scalability of \name Solver} 
\begin{figure}[!t]
    \includegraphics[width=1.\linewidth]{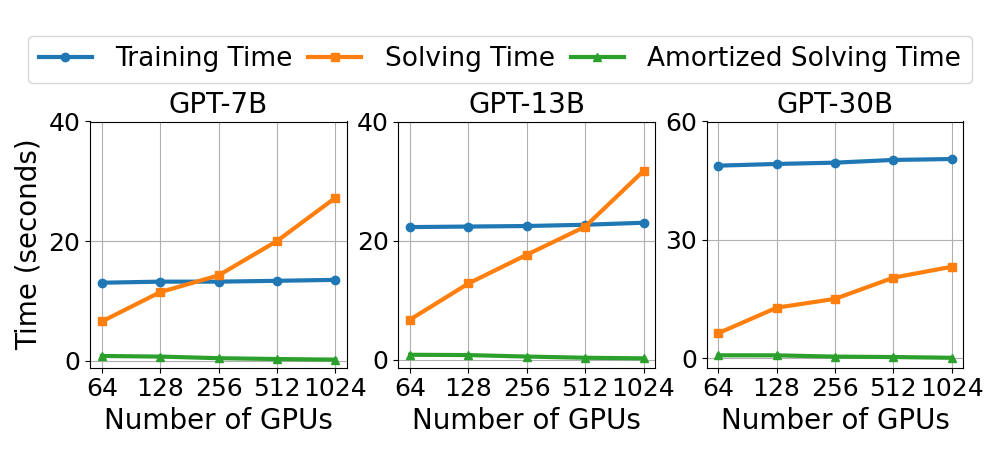}
    \caption{
    Estimated per-iteration training time, solving time and amortized solving time (solving time divided by the number of GPU nodes, i.e., $N/8$) of \name Solver.
    }
    \label{fig:solver_scaling}
\end{figure}

We demonstrate that the MILP problem solving in \name solver scales well and can be fully overlapped by the training. 
Due to the sequence bucketing, the MILP solving time (Eq.~\eqref{eq:minimize_time_bkt}) is mainly affected by the number of GPUs ($N$) and the number of buckets. 
Since the number of buckets is small, we focus on the scalability against $N$. 
When there are more GPUs, the problem solving for each iteration takes longer, while the training time maintains at a similar level (as it is common to scale the batch size proportionally to $N$). 
However, recall that \name disaggregates the problem solving on CPUs and the training on GPUs (\S\ref{sec:imple}). 
In modern clusters, using more GPUs also indicates more CPUs are available to solve the MILP for different batches concurrently, so the solving time can be amortized and we can still overlap the problem solving well.
To elaborate, we conduct a simulation experiment in Fig.~\ref{fig:solver_scaling} to assess the time cost with GPU counts varying from 64 to 1024. 
We show the estimated training time, solving time and amortized solving time below (solving time divided by the number of GPU nodes, i.e., $N/8$) per iteration in seconds.
We find that the estimated training time does remain similar.
Although the per-iteration solving time may exceed training time when $N\geq$ 256, the amortized solving time is much lower, and is fully overlappable.

\subsubsection*{More Experimental Results.} 
Due to the space limitation, we put more results and analysis in the \ifasplos\else{Appendix}\fi, including more details of our experimental setups in Appendix~\ifasplos\else{\ref{sec:appendix_exp_setup}}\fi, estimation accuracy of cost models in Appendix~\ifasplos\else{\ref{sec:appendix_est_acc}}\fi, performance analysis of SOTA systems in Appendix~\ifasplos\else{\ref{sec:appendix_ana_sota}}\fi, discussion about integrating context parallelism in Appendix~\ifasplos\else{\ref{sec:appendix_integrate_cp}}\fi.
Please refer to our Appendix for more details.

\section{Related Work} \label{sec:related}

\subsubsection*{Parallel Training Optimization}
During hybrid parallel training, the optimization of parallel strategies is crucial.
Many advanced parallelism optimization techniques~\cite{DBLP:conf/mlsys/flexflow,unger2022unity,DBLP:ALPA,DBLP:Galvatron,Galvatron-BMW,JCST:AI-system-review} are developed to automate the tuning of parallel strategies.
However, these works are designed mainly for homogeneous training corpora, while \name focuses on flexible strategies for data with heterogeneous lengths.



\subsubsection*{Long Context Training}

Efforts to optimize long context training have led to various elaborate parallel strategies, such as ring attention for LLMs~\cite{DBLP:RSA,DBLP:ring-attn,distflashattn,DBLP:striped_attn}, though they often suffer from communication overhead and inefficiencies with severe communication cost.
These methods are orthogonal, and can be integrated into \name seamlessly, detailed in \ifasplos\else{Appendix~\ref{sec:appendix_integrate_cp}.}\fi
Other works aim to support long context training by extending attention mechanism~\cite{DBLP:conf/acl/DaiYYCLS19, DBLP:journals/corr/abs-2004-05150} and optimizing position embedding ~\cite{ding2024longropeextendingllmcontext, zhu2024longembedextendingembeddingmodels}, which are also orthogonal.

\subsubsection*{Heterogeneous Cluster Training}
Training efficiency on heterogeneous GPU clusters is the main topic of these works~\cite{DBLP:HetPipe,DBLP:AccPar,DBLP:AMP,whale-atc22}, focusing on the heterogeneity of hardware.
In contrast, \name emphasizes flexible parallelism to address the workload heterogeneity of varied-length data.


\section{Conclusion} \label{sec:conclusion}
In this paper, we propose an efficient system to accelerate LLM training via flexible and adaptive sequence parallelism.
Specifically, \name addresses the workload heterogeneity and optimizes the parallelism strategies for varied-length training corpora in real-world datasets.
Experiments demonstrate that \name outperforms SOTA systems by up to 1.98$\times$.

\begin{acks}
This work is supported by National Science and Technology Major Project (2022ZD0116315), National Natural Science Foundation of China (U23B2048, 62402011), Beijing Municipal Science and Technology Project (Z231100010323002), China National Postdoctoral Program for Innovative Talents (BX20230012), China Postdoctoral Science Foundation (2024M750103), Beijing Natural Science Foundation (4244080), research grant No. IPT-2024JK29, ByteDance-PKU joint program, and High-performance Computing Platform of Peking University. Fangcheng Fu and Bin Cui are the corresponding authors.
\end{acks}


\bibliographystyle{ACM-Reference-Format}
\balance
\bibliography{my-references}

\ifasplos
\else
\appendix

\clearpage
\section{Momery-balanced Micro-batch Chunking in Sequence Blaster} \label{sec:appendix_mb_chunk}
We illustrate the memory-balanced micro-batch chunking algorithm in Sequence Blaster based on dynamic programming.
Specifically, given a batch of sequences $\mathcal{B}=\{\mathcal{S}_k\}$ that has already been sorted according to takeaway \#2, we split them into consecutive $M$ micro-batches, and micro-batch $\mathcal{M}_i$ contains all sequences $k$ satisfying $j_{i-1} \leq k < j_i$, where $j_i$ is the ending indices for $\mathcal{M}_i$ ($j_0=0$).
To balance the token amount of each micro-batch, we aim to minimize the maximum total token number of each micro-batch as follows:
{
\begin{equation}
{
\small
    \argmin_{\{j_i\}} \max_{i \in [1,M]}\{\sum_{k \in [j_{i-1}, j_i)}s_k\}.
}
\end{equation}
}
Again, we solve the problem via dynamic programming. 
Denote $DP[k][i]$ as the optimal value when blasting the first $k$ sequences into $i$ micro-batches. 
Starting with $DP[0][0]=0$, we can solve the problem via the following state transition formula:
{
\begin{equation}
{
\small
    DP[k][i] = \min_{j \in [i-1,k-1]} \{\max\{ DP[j][i-1], \sum_{l \in [j+1,k]}s_l \} \},
}
\end{equation}
}
where $\sum_{l \in [j+1,k]}s_l$ represents the total token number of the $i^{th}$ micro-batch when splitting micro-batch at the $\mathcal{S}_j$. $DP[K][M]$ denotes the optimal solution, and the optimized values of $j_i$ splits the global batch data into $M$ micro-batches with balanced memory consumption.

\section{Details of Experimental Setups} 
\label{sec:appendix_exp_setup}
\subsection{Model Configuration}
\label{sec:appendix_model_config}
Tab. ~\ref{tab:model_config} presents the specific configurations of the GPT-7B, 13B and 30B models used in our experiments.
The parameter number of each model in Tab.~\ref{tab:model_config} is under maximum context length of 384K, where the positional embedding contains 1-2 billion parameters.

\begin{table}[H]
    \centering
    \caption{
        Model configuration (384K max context length).
    }
    \label{tab:model_config}
    \begin{tabular}{c|c|c|c}
    \toprule
    Model   & \# Layers & \# Param  & Hidden Dim \\ \hline\hline
    GPT-7B  & 32        & 7.85B     & 4096       \\ \hline
    GPT-13B & 40        & 14.03B    & 5120       \\ \hline
    GPT-30B & 60        & 32.72B    & 6656       \\
    \bottomrule
    \end{tabular}
\end{table}

\subsection{Protocols}
\label{sec:appendix_parallel_config}
For fair comparison, we manually tune the most efficient parallelism strategies for all baseline systems under different workloads.
For DeepSpeed, the optimal strategy is usually among SP=64 or SP=32 with ZeRO-3, and for Megatron-LM, the optimal strategy is usually among TP=8, CP=8, or TP=16, CP=4, or TP=8, CP=4, DP=2 with ZeRO-1.

We also apply activation checkpointing strategies for each system to make sure all systems can fit the models without out-of-memory issues.
For GPT-7B, activation checkpointing is unnecessary to support a 384K context length on 64 GPUs.
For GPT-13B, we only checkpoint MLP layers, while for GPT-30B, almost all layers are checkpointed to support 384K context length.


\section{Estimation Accuracy of Cost Models} 
\label{sec:appendix_est_acc}

We evaluate the accuracy of the cost estimator (\S\ref{subsubsec:cost_est}) utilized in \name across diverse configurations (as shown in Tab.~\ref{tab:obs}), including sequence parallelism degree, batch size, and sequence length.
Fig.~\ref{fig:estimation-acc} compares the deviation of the estimated cost and the empirical execution time.
As can be seen, our overhead estimator adeptly approximates the execution overhead, with discrepancies consistently remaining below 5\%.
The accurate estimations rendered by the estimator facilitates performance of our system.

\begin{figure}[H]
    \includegraphics[width=.96\linewidth]{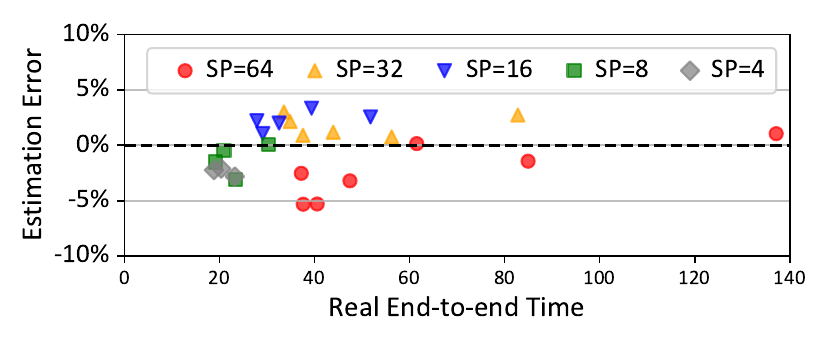}
    \caption{Estimation accuracy.}
    \label{fig:estimation-acc}
\end{figure}

\section{Performance Analysis of SOTA Systems} 
\label{sec:appendix_ana_sota}
Here we analyze the performance of SOTA systems, i.e., DeepSpeed and Megatron-LM.
As shown in Fig. ~\ref{fig:main-result}, in most cases, DeepSpeed has similar or better performance than Megatron-LM.
This is attributed to the different mechanism of CP in Megatron-LM and SP in DeepSpeed, as well as the skewness of datasets.
CP usually has much more communication volumn than the All-to-All in SP.
Although CP leverages the overlap between attention computation and KV transmission to hide the communication overhead, in scenarios with limited inter-node bandwidth and a majority of short sequences in datasets, the attention computation often fails to hide the communication.
Therefore, Megatron-LM’s performance is usually constrained by its higher communication volume compared to DeepSpeed and \name.

\section{Integrating Context Parallelism} 
\label{sec:appendix_integrate_cp}
Context parallelism~\cite{DBLP:RSA,DBLP:ring-attn,distflashattn,DBLP:striped_attn} is also an important technique for long context training, as illustrated in~\S\ref{other_parallel}.
As a different paradigm to scatter sequences across devices, CP is orthogonal to our work, and can be integrated into \name.
Specifically, similar to \name utilizing ZeRO and flexible SP, we can employ TP, ZeRO and CP, fix the parallelism degree of TP, and employ the flexible sequence parallelism strategy of \name to achieve flexible CP, which adaptively adjusts the CP group size according to the varied-length sequences.
We'll integrate CP into our system as the future work.

\fi

\end{document}